\documentclass[final,twocolumn,pra,showpacs,preprintnumbers,floatfix]{revtex4}
\pdfoutput=1

\usepackage{graphicx}
\usepackage{amsmath,amssymb}

\newcommand{\assign}{:=}

\newcommand{\mathi}{\mathrm{i}}
\newcommand{\mathe}{\mathrm{e}}

\newcommand{\tmtextbf}[1]{{\bfseries{#1}}}
\newcommand{\tmtextit}[1]{{\itshape{#1}}}

\newcommand{\tmem}[1]{{\em #1\/}}

\newcommand{\tmop}[1]{\ensuremath{\text{#1}}}

\begin{document}

\title{Nonclassical correlations from dissociation time entanglement}

\author{Clemens Gneiting}
\author{Klaus Hornberger}
\affiliation{Arnold Sommerfeld Center for Theoretical Physics, 
Ludwig-Maximilians-Universit{\"a}t M{\"u}nchen, Theresienstra{\ss}e 37, 
80333 Munich, Germany
}

\preprint{published in: Appl. Phys. B 95 (2009) 237-244}

\begin{abstract}
We discuss a strongly entangled two-particle state of motion that emerges naturally
from the double-pulse dissociation of a diatomic molecule. This state, which may be
called dissociation-time entangled, permits the unambiguous demonstration
of nonclassical correlations by violating a Bell inequality based on
switched single-particle interferometry and only position measurements.
We apply time-dependent scattering theory to determine the detrimental effect of
dispersion. The proposed setup brings into reach the possibility of establishing
nonclassical correlations with respect to system properties that are truly
macroscopically distinct.
\end{abstract}



\maketitle

\section{Introduction}

Quantum mechanics (QM) has been forcing physicists to think about its relation
to our classical perception of the world ever since its formulation. Probably
its most unsettling feature is the rejection of the notion of the trajectory
of a proper, material particle. The at every instant well-defined, definite
position of a point particle in space (or rather phase space), as we deduce it
from our everyday experience of the macroscopic world, is probably the most
basic assumption in classical physics; it might rightly be called its core.
Quantum mechanics, on the other hand, permits superpositions of arbitrarily
remote positions of a particle. The second cornerstone of classical physics,
locality, is equally abandoned in QM by entanglement, which entails
correlations between spatially separated parties that cannot be reduced to
shared randomness
{\cite{Einstein1935a,Schrodinger1935a,Bell1964a,Aspect1982a}}. So what could
be a more striking demonstration of the failure of classical concepts than the
experimental generation of a state that combines both, particles
simultaneously being at macroscopically distinct positions and being
entangled in these positions? In this article, we would like to promote such a
state, whose experimental demonstration seems to be in reach of present day
technology.

\begin{figure}[tb]
  \includegraphics[width=\columnwidth]{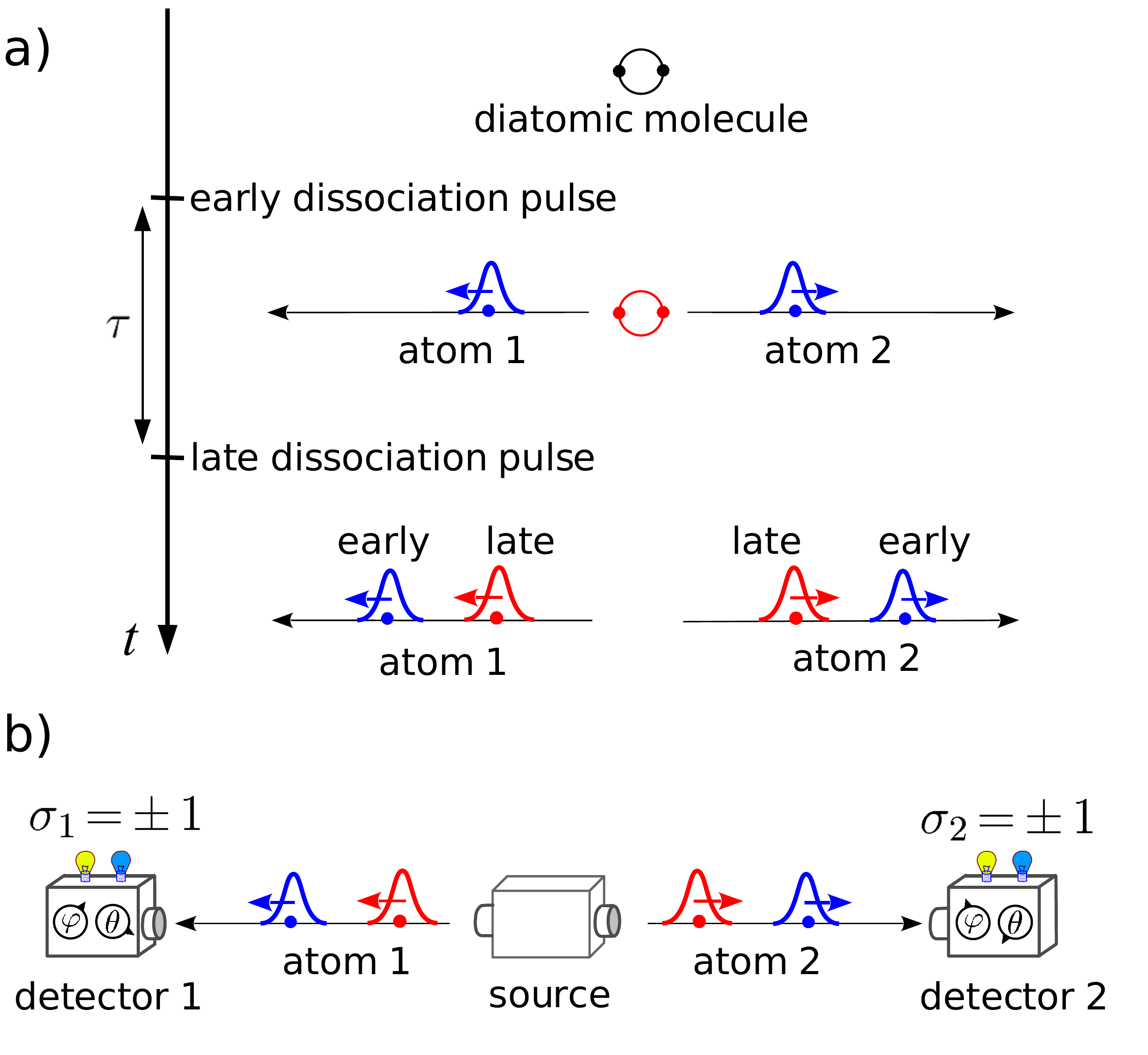}
  \caption{\label{DTEGeneration}(color online) (a) Generation of a
  dissociation-time entangled pair of atoms. A diatomic molecule is exposed to
  two subsequent dissociation pulses separated by a time period $\tau$. The
  early pulse is chosen such that the atoms remain bound with a given
  probability. The second pulse dissociates the remaining molecular state
  component. Each atom then has an early and a late wave packet component such
  that the resulting state of two counterpropagating particles is entangled in
  the dissociation times. (b) If the wave packet components of the
  dissociation-time entangled atoms are spatially sufficiently separated, they
  constitute effectively dichotomic properties which are amenable to a Bell
  test in terms of switched single-particle interferometry and position
  measurements, see Fig. \ref{MachZehnder}.}
\end{figure}

The approach is to trigger the dissociation of a diatomic molecule coherently
at two different instants (``early'' and ``late'') separated by a time period
$\tau$, as depicted in Fig. \ref{DTEGeneration}(a). The dissociation is
coherent in the sense that no information is leaked to the environment about
the time of dissociation. Each atom then has an early and a late wave packet
component such that the resulting state of two counterpropagating particles is
entangled in the dissociation times. For this reason, such a state might be
called ``dissociation-time entangled'' (DTE). A macroscopic time period $\tau$
would result in a superposition of truly macroscopically distinct wave
packets. Experimental setups based on ultracold Feshbach molecules indeed
render such a scenario realistic {\cite{Gneiting2008a}}.

Of course, generating a highly nonclassical state is only half of the job to
be done. Just as important is the verification of its nonclassical nature.
Here a subtlety comes into play: Whom do we want to convince? Do we face a
person that believes in QM, or do we face a cocksure classical physicist, who
understands how to measure a particle position, but knows nothing about
quantum physics? Several proposals for demonstrating nonclassicality in the
motion of material particles are content with the confirmation of the
entangled nature of the underlying quantum states, while suffering from the
caveat that they have to presume the validity of QM in order to be convincing.
From an operational point of view, they do not imply nonclassicality by
establishing nonlocal correlations. They therefore fail to convince the
cocksure classical physicist---or an unprejudiced layman. This holds in
particular for proposals that are close in spirit to the original
Einstein--Podolsky--Rosen (EPR) argument
{\cite{Einstein1935a,Kheruntsyan2005a,Opatrny2001a}}, since the positive
Wigner function of the EPR state admits a classical interpretation. One
possibility to overcome this would be to consider observables which have no
classical analogue, such as the displaced parity or pseudospin operators, with
nonpositive phase space representations {\cite{Banaszek1998a,Chen2002a}}. However,
their experimental implementation seems to be exceedingly difficult in the
case of free material particles, where only position measurements are easily
realized.

The DTE state differs fundamentally from the Gaussian states describing
EPR-type correlations. As opposed to Gaussian states, it displays a strongly
structured, partly negative Wigner function and it is amenable to an
unambiguous demonstration of nonclassicality in terms of simple position
measurements. This is achieved by implementing a Bell experiment, as will be
elaborated below, see Fig. 1(b). Within a range of tolerance as imposed by dispersion, the
resulting nonlocal correlations would manifest both the coherence between the
macroscopically distinct early and late wave packet components and their
entanglement. The proposed setup would therefore permit the demonstration of
nonclassicality in the motion of material particles on macroscopic scales,
making quantum mechanical counterintuitiveness concrete for anyone, even the
layman.

The article is structured as follows: In Section \ref{TimeBinEntanglement} we
recall the concept of time-bin entanglement, which was originally applied to
photons {\cite{Brendel1999a,Tittel2000a,Simon2005a}}. We carry it over to
material particles and discuss the implications for a Bell test based on
interferometric state transformation and subsequent detection. In particular,
we show why the dispersion-induced spreading of the wave packets, which is
absent in the photonic case, does not diminish the ensuing nonlocal
correlations. In Section \ref{DissociationTimeEntanglement} we investigate the
analogue Bell test based on dissociation-time entanglement, which corresponds
to the more natural state generation scenario. The dispersion-induced
distortion between the early and the late wave packets is shown to affect the
nonlocal correlations. We give benchmark criteria for the capability of
establishing nonclassicality. In Section \ref{Conclusions} we finally
summarize the advantages of a demonstration of nonclassicality based on a DTE
Bell test, which includes in particular the possibility to go to macroscopic
scales.

\section{\label{TimeBinEntanglement}Time-bin entanglement}

\begin{figure}[tb]
  \includegraphics[width=\columnwidth]{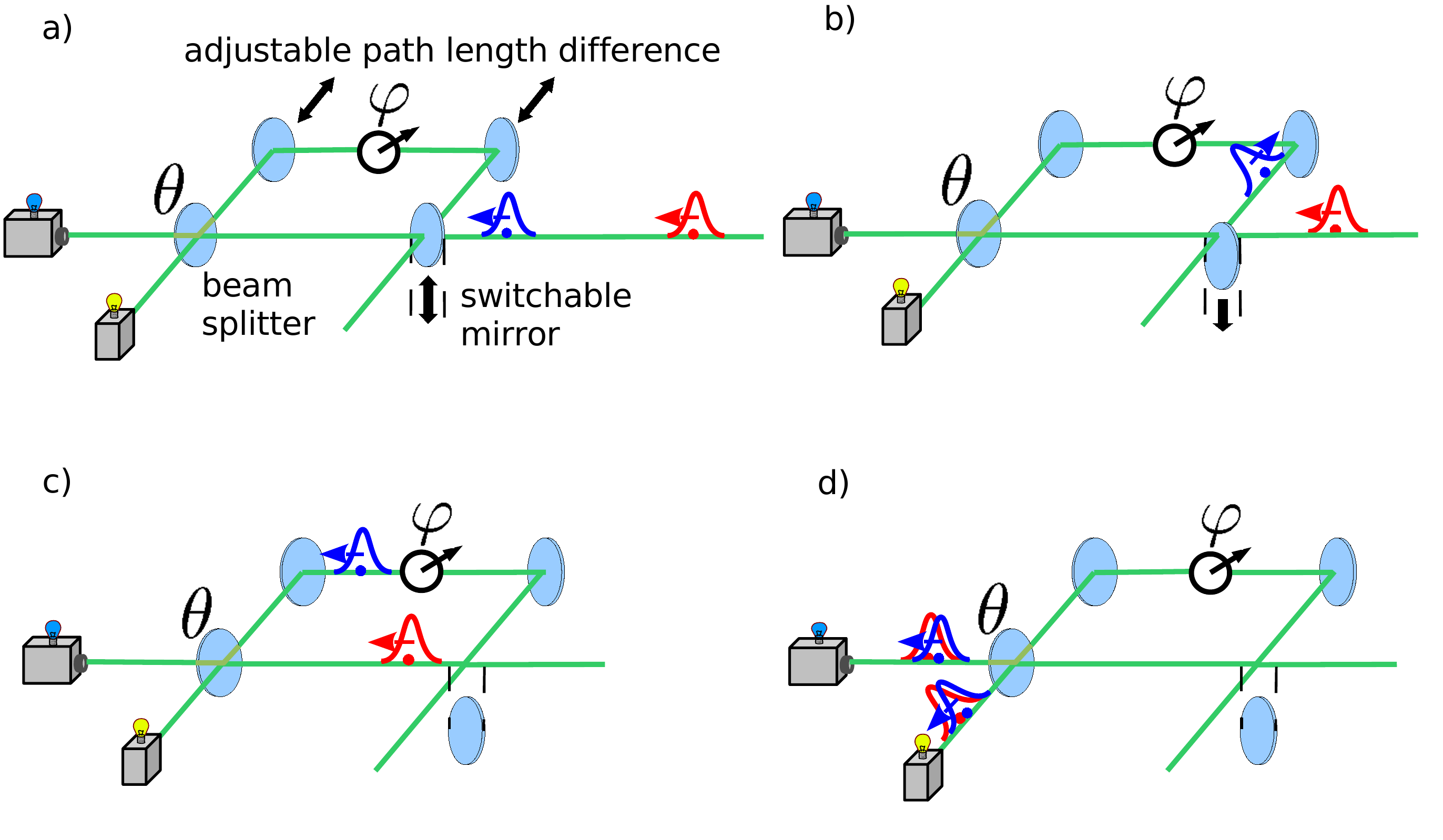}
  \caption{\label{MachZehnder}(color online) State analysis with an asymmetric
  Mach--Zehnder interferometer. (a) When the early wave packet component
  arrives, the switchable mirror is in place and deflects it into the long
  arm. (b) The switchable mirror is removed before the arrival of the late wave
  packet. (c) The early wave packet acquires an optional phase shift $\varphi$
  in the long arm, whereas the late wave packet propagates through the short
  arm. (d) The path difference is chosen such that it cancels the distance
  between the early and the late wave packet component. Detecting the atom in
  one of the output ports, at a given phase $\varphi$ and splitting ratio
  $\theta$, amounts to a measurement analogous to a spin-1/2 detection in an
  arbitrary direction.}
\end{figure}

The idea to encode qubits in spatially distinct wave packets was introduced in
{\cite{Brendel1999a}}, building upon the concept of energy-time entanglement
{\cite{Franson1989a}}. A pulsed laser in combination with an asymmetric
interferometer placed in front of a parametric down-conversion crystal permits
to generate twin-photons which are entangled in their creation time. The
resulting state, which may be called time-bin entangled (TBE), takes the form
\begin{equation}
  \left. \label{TBEstate} | \Psi_{\tmop{tbe}} \right\rangle =
  \frac{1}{\sqrt{2}} (| \text{E} \rangle_1 | \text{E} \rangle_2 +
  \mathe^{\mathi \phi_{\tau}} | \text{L} \rangle_1 | \text{L} \rangle_2),
\end{equation}
where $| \text{E} \rangle_i$ and $| \text{L} \rangle_i$ denote spatially
distinct traveling photonic modes corresponding to the early and the late
creation time. The TBE state (\ref{TBEstate}) can be visualized analogously to
the DTE state depicted in Fig.\,\ref{DTEGeneration}(b). Since the relevant
entanglement resides in the relation between these modes, we can interpret the
creation times as a dichotomic property constituting an effectively
two-dimensional state space (per photon). By identifying, say, ``early'' with
``spin up'' and ``late'' with ``spin down'', the state $| \Psi_{\tmop{tbe}} \rangle$
evidently corresponds to the Bell state
\begin{equation}
  \left. \label{BellState} | \Psi_{\tmop{spin}} \right\rangle =
  \frac{1}{\sqrt{2}} (| \uparrow \rangle_1 | \uparrow \rangle_2 +
  \mathe^{\mathi \phi} | \downarrow \rangle_1 | \downarrow \rangle_2) .
\end{equation}
Provided that the early and the late wave packet components are spatially
sufficiently distinct, a switched, asymmetric Mach--Zehnder interferometer
makes it possible to perform the analogue of a general spin measurement, see
Fig.\,\ref{MachZehnder} and the discussion below {\cite{Brendel1999a}}. Such a
TBE state has indeed been used successfully with photons, e.g., for
establishing nonlocal correlations over fiber distances of more than 50 km
{\cite{Brendel1999a,Tittel2000a,Simon2005a}} in a similar setup as in
Fig.\,\ref{MachZehnder}, though without switching.

The matter wave analogue of the TBE state (\ref{TBEstate}) follows from
identifying $| \text{E} \rangle_i$ and $| \text{L} \rangle_i$ with the early
and the late wave packet components of two freely moving atoms. It should be
emphasized, though, that when it comes to material particles, the TBE state
(\ref{TBEstate}) is not the natural outcome in a two-time dissociation
process. One conceivable scenario relies on the controlled dissociation of a
weakly bound Feshbach molecule with the help of a Feshbach resonance
{\cite{Gneiting2008a}}. An appropriately chosen magnetic field pulse causes
the atomic components to dissociate and propagate in opposite directions
{\cite{Mukaiyama2004a,Durr2004b}}. A sequence of two magnetic field pulses
then generates the desired superposition of two dissociation times. A generic
dissociation-time entangled (DTE) state thus takes the form
\begin{equation}
  \label{DTEstate} | \Psi_{\tmop{dte}} \rangle = \frac{1}{\sqrt{2}}  \left(
  \widehat{\text{U}}^{(0)}_{\tau} | \Psi_0 \rangle + \mathe^{\mathi
  \phi_{\tau}} | \Psi_0 \rangle \right)
\end{equation}
with $| \Psi_0 \rangle = \frac{1}{\sqrt{2}} | \psi^{\tmop{cm}}_0 \rangle
\left( | \psi^{\tmop{rel}}_0 \rangle + \widehat{\text{P}} |
\psi^{\tmop{rel}}_0 \rangle \right)$. Here, $\widehat{\text{P}}$ is the parity
operator, and we assume that the early and the late dissociation occur with
equal probability. In the following, we take the state $| \Psi_{\tmop{dte}}
\rangle$, and accordingly $| \Psi_{\tmop{tbe}} \rangle$, to describe a
one-dimensional, longitudinal motion, implying that the transverse components
of the motion are confined to the ground state of an atom guide. The state $|
\psi^{\tmop{rel}}_0 \rangle$ denotes the wave packet of the relative motion
after a single dissociation pulse propagating into positive direction, while
$| \psi^{\tmop{cm}}_0 \rangle$ describes the wave packet of the center of mass
motion resting at the former position of the molecule. The DTE state
(\ref{DTEstate}) differs from the above TBE state (\ref{TBEstate}) as it is
not composed of two single-particle product states. Rather, it superposes the
relative coordinate of the two atoms. It also incorporates the unavoidable
dispersion-induced distortion between the early and the late state components,
described by the free time-evolution operator
$\widehat{\text{U}}^{(0)}_{\tau}$.

Both the TBE state (\ref{TBEstate}) and the DTE state (\ref{DTEstate}) are
amenable to a Bell test based on interferometric state transformation and
subsequent position measurement. Before tackling the motional Bell experiment
with the experimentally appropriate DTE state (\ref{DTEstate}), it is
instructive to first describe the interferometric setup and the effect of
dispersion with the more transparent TBE state (\ref{TBEstate}). The setup is
based on switched, asymmetric Mach--Zehnder interferometers, as shown in Fig.
\ref{MachZehnder}. Their action can be described as follows: The early wave
packet components are deflected by the switches into the long arms of the
interferometers, where they acquire optional phase shifts $\varphi_i$ before
the beam splitters distribute them onto their output ports according to their
splitting ratios, as parameterized by the angles $\theta_i$. The switches are
timed such that they let pass the late wave packet components, which then
propagate straight to the beam splitters and are also distributed according to
the splitting ratios. When the path length difference between the two
interferometer arms is chosen such that it exactly cancels the distance
between the early and the late wave packets and given ideal phase shifters and
beam splitters, the early and late state components $| \text{E} \rangle_i$ and
$| \text{L} \rangle_i$ transform according to
\begin{eqnarray}
  | \text{E} \rangle_i & \rightarrow & \mathe^{\mathi \varphi_i} \cos \theta_i
  | + \rangle_i + \mathe^{\mathi \varphi_i} \sin \theta_i | - \rangle_i 
  \label{TBEtransformation}\\
  | \text{L} \rangle_i & \rightarrow & \sin \theta_i | + \rangle_i - \cos
  \theta_i | - \rangle_i, \nonumber
\end{eqnarray}
where $| + \rangle_i$ and $| - \rangle_i$ denote wave packet components in the
two output ports of the $i$th interferometer at a stage of the time evolution
when the particles already passed the interferometers. This mapping assumes
that the components $| \text{E} \rangle_i$ and $| \text{L} \rangle_i$ of the
initial state are identical wave packets up to their spatial displacement: $|
\text{L} \rangle_i = \mathe^{\mathi \widehat{\text{p}}_i s_i / \hbar} |
\text{E} \rangle_i$, with $s_i$ denoting their separation. Moreover, the
states $| \text{E} \rangle_i$ and $| \text{L} \rangle_i$ must be sufficiently
spatially distinct, such that they can be distinguished by the switches, even
when taking dispersion-induced wave packet spreading into account. Note that
it is not important to be specific about the particular instant, at which we
consider the wave packet components $| + \rangle_i$ and $| - \rangle_i$, since
we will only be interested in the overall detection probabilities per port,
which unitary time evolution guarantees to remain unaffected by any subsequent
time evolution. In this sense, the present treatment already incorporates
dispersion, since we do not have to assume that the wave packets $| +
\rangle_i$ and $| - \rangle_i$ are equal in shape to the wave packets $|
\text{E} \rangle_i$ and $| \text{L} \rangle_i$.

After passing the interferometers, the time-bin entangled state can be written
in accordance with (\ref{TBEtransformation}) as
\begin{equation}
  | \Psi'_{\tmop{tbe}} \rangle = \frac{1}{2^{2 / 3}}  \sum_{\sigma_1, \sigma_2
  = \pm} \left[ \mathe^{\mathi (\varphi_1 + \varphi_2)} + \sigma_1 \sigma_2
  \mathe^{\mathi \phi_{\tau}} \right] | \sigma_1 \rangle_1 | \sigma_2
  \rangle_2 .
\end{equation}
For clarity, we have taken 50:50 beam splitters $(\theta_1 = \theta_2 = \pi /
4)$. As we will see, this restriction does not come with any loss of
significance, since it still allows for a maximal violation of a Bell
inequality. The joint probability for detecting the particles in a particular
output port combination thus reads $(\sigma_{1, 2} = \pm 1)$
\begin{eqnarray}
  \text{P}_{\tmop{tbe}} (\sigma_1, \sigma_2 | \varphi_1, \varphi_2) & = & |
  \langle \sigma_1 |_1 \langle \sigma_2 |_2 | \Psi'_{\tmop{tbe}} \rangle |^2 
  \label{TBEdetectionProbability}\\
  & = &  \frac{1}{4} [1 + \sigma_1 \sigma_2 \cos (\varphi_1 + \varphi_2 -
  \phi_{\tau})] . \nonumber
\end{eqnarray}
From an experimental point of view, the natural quantity to be measured is the
arrival time of the particles at the detectors. The two-time probability
density $\tmop{pr} (\sigma_1, \sigma_2 ; t_1, t_2)$ for detecting the
particles at times $t_1$, $t_2$ in a particular output port combination
constitutes in general a complicated fringe pattern as a function of the
$\varphi_i$, $\theta_i$ and $t_i$. It depends not only on the shape of the
wave packets and their dispersion-induced modifications due to the overall
propagation time, but also on the particular implementation of the measurement
{\cite{Allcock1969a,Werner1986a,Aharonov1998a,Muga2000a}}. The integrated
probability $\text{P}_{\sigma_1, \sigma_2} = \int  \text{d} t_1  \text{d} t_2 
\text{$\tmop{pr} (\sigma_1, \sigma_2 ; t_1, t_2)$}$, however, only measures
the overall likelihood for finding the particles in a particular output port
combination. It is thus unaffected by the particular shape of the wave packets
and any dispersion-induced modification thereof, since unitary time evolution
implies the conservation of probability in the output ports. Strictly
speaking, this is valid only if measurement-induced reflections of the wave
packets at the detectors can be excluded
{\cite{Allcock1969a,Werner1986a,Aharonov1998a,Muga2000a}}. But since our setup
does not require prominent temporal or spatial resolution of the measurement,
we can safely neglect this effect. For this reason, we may identify the
integrated probability $\text{P}_{\sigma_1, \sigma_2}$ with
(\ref{TBEdetectionProbability}). From the joint probability
(\ref{TBEdetectionProbability}) we easily obtain the corresponding correlation
function:
\begin{eqnarray}
  C_{\tmop{tbe}} (\varphi_1, \varphi_2) & = & \sum_{\sigma_1 \sigma_2 = \pm}
  \sigma_1 \sigma_2  \text{P}_{\tmop{tbe}} (\sigma_1, \sigma_2 | \varphi_1,
  \varphi_2) \nonumber\\
  & = & \cos (\varphi_1 + \varphi_2 - \phi),  \label{TBEcorrelationFunction}
\end{eqnarray}
which violates the CHSH-inequality {\cite{Clauser1969a}} $|C (\varphi_1,
\varphi_2) + C (\varphi_1, \varphi'_2) + C (\varphi'_1, \varphi_2) - C
(\varphi'_1, \varphi'_2) | \leqslant 2$ maximally, e.g., for the choices
$\varphi_1 = \phi / 2$, $\varphi_2 = \phi / 2 - \pi / 4$, $\varphi'_1 = \phi /
2 + \pi / 2$, and $\varphi'_2 = \phi / 2 + \pi / 4$.

Like for the (switched) photonic TBE Bell setup, the TBE correlation function
for the material particles (\ref{TBEcorrelationFunction}) is identical to the
correlation function obtained in the spin-based Bell experiment with the state
(\ref{BellState}). We thus find that the nonlocal correlations due to the TBE
state (\ref{TBEstate}) are not affected by dispersion. This follows from the
assumption that the early and the late wave packets are identical up to a
spatial displacement. We will see below that this is not valid for the more
realistic DTE state (\ref{DTEstate}), where the dispersion-induced distortion
between the early and the late wave packets is taken into account. We note
that the photonic experiments performed to date have been done without switch,
using instead a postselection procedure {\cite{Brendel1999a}}. This has been
shown to allow for a local hidden variable model
{\cite{Aerts1999a,Cabello2008a}}, a drawback that is avoided if the switching
can be easily implemented, as is the case with slow material particles.

In order to make the correspondence with the spin-based Bell experiment most
transparent, we conclude this section by recalling that the above state
transformation (\ref{TBEtransformation}) acting on a general (single-particle)
``time-bin'' superposition state $| \psi \rangle = a | \text{E} \rangle + b |
\text{L} \rangle,$ $|a|^2 + |b|^2 = 1$, followed by the detection in one of
the output ports can formally be understood as the analogue of a ``spin''
measurement with respect to the measurement axis $\vec{n} = (\sin (2 \theta)
\cos \varphi, \sin (2 \theta) \sin \varphi, \cos (2 \theta))$. To this end, we
define the analogues of Pauli matrices
\begin{eqnarray*}
  \hat{\sigma}_x & = & | \text{E} \rangle \langle \text{L} | + | \text{L}
  \rangle \langle \text{E} |\\
  \hat{\sigma}_y & = & - i | \text{E} \rangle \langle \text{L} | + i |
  \text{L} \rangle \langle \text{E} |\\
  \hat{\sigma}_z & = & | \text{E} \rangle \langle \text{E} | - | \text{L}
  \rangle \langle \text{L} |,
\end{eqnarray*}
and $\widehat{\vec{\sigma}} = ( \hat{\sigma}_x, \hat{\sigma}_y,
\hat{\sigma}_z)$. The interferometric transformation (\ref{TBEtransformation})
and the subsequent detection of the transformed state $| \psi' \rangle$ in the
output ports can then be equivalently understood as a measurement of the
observable $\vec{n} \cdot \widehat{\vec{\sigma}}$ with the untransformed state
$| \psi \rangle$. This is most easily seen by noting that the transformation
(\ref{TBEtransformation}) acts like an expansion of the incoming time-bin
state into the eigenstates $| \vec{n}, + \rangle, | \vec{n}, - \rangle$ of
$\vec{n} \cdot \widehat{\vec{\sigma}}$. With this, the joint detection
probability may equally well be written as
\[ \text{P}_{\tmop{tbe}} (\sigma_1, \sigma_2 | \vec{n}_1, \vec{n}_2) = |
   \langle \vec{n}_1, \sigma_1 |_1 \langle \vec{n}_2, \sigma_2 |_2 |
   \Psi_{\tmop{tbe}} \rangle |^2, \]
and the correlation function is given by
\[ C_{\tmop{tbe}} ( \vec{n}_1, \vec{n}_2) = \langle \Psi_{\tmop{tbe}} | \left[
   \vec{n}_1 \cdot \widehat{\vec{\sigma}}_1 \right] \otimes \left[ \vec{n}_2
   \cdot \widehat{\vec{\sigma}}_2 \right] | \Psi_{\tmop{tbe}} \rangle . \]
This correspondence implies that the above time-bin scenario can be used to
apply other quantum information strategies to the motion of material
particles, such as the teleportation of a qubit state. Encoding qubits in
spatially distinct wave packets thus permits general single-qubit state
processing to be based only on matter wave optics and subsequent position
measurements. This gives us a simple and robust method at hand for performing
the most relevant quantum tests in the {\tmem{motion}} of material particles.

\section{\label{DissociationTimeEntanglement}Dissociation-time entanglement}

Up to now, we mainly investigated the Bell test for the TBE state
(\ref{TBEstate}). However, from an experimental point of view, the DTE state
(\ref{DTEstate}) is much more relevant, since its structure is naturally
produced by the dissociation of a diatomic molecule. The DTE wave function
differs from the TBE state in two important points: (i) it does not separate
into single-particle states at a particular dissociation time and (ii) the
dispersive time evolution between the early and the late dissociation process
implies different shapes for the early and late wave packet components.
Furthermore, we now specify the phase shifters to be implemented by varying
the arm lengths of the interferometers, which effects an additional mismatch
between the early and the late state components. All these modifications
require a more sophisticated theoretical description of the setup. It turns
out that time-dependent scattering theory provides the appropriate framework.

Scattering theory applies when the system dynamics under consideration permits
to relate asymptotic {\tmem{in}}-states to asymptotic {\tmem{out}}-states
{\cite{Taylor1972a}}. Then, given one is only interested in the relation
between these asymptotic states, the exact time evolution connecting the
{\tmem{in}} and the {\tmem{out}} states can be split up into an
``instantaneous'' transformation describing the accumulated effect of the
interaction potential and a subsequent free time evolution. In our case,
\[ | \Psi'_{\tmop{dte}} \rangle \assign \widehat{\text{U}}_t |
   \Psi_{\tmop{dte}} \rangle = \widehat{\text{U}}^{(0)}_t  \widehat{\text{S}}
   | \Psi_{\tmop{dte}} \rangle, \]
where $| \Psi'_{\tmop{tbe}} \rangle$ denotes the TBE state at a given stage of
evolution when the passage through the interferometers is completed, and the
scattering operator $\widehat{\text{S}}$ describes the ``raw'' action of the
interferometers. In our case, another subtlety comes into play, since we have
to distinguish the effect of the interferometers depending on whether the
switch is in place (``on''), enforcing deflection into the longer arm, as for
the early state component, or absent (``off''), admitting undeflected passage,
as for the late state component:
\[ | \Psi'_{\tmop{dte}} \rangle = \widehat{\text{U}}^{(0)}_t  \left(
   \widehat{\text{S}}^{(\tmop{on})}  \widehat{\text{U}}^{(0)}_{\tau} | \Psi_0
   \rangle + \mathe^{\mathi \phi_{\tau}}  \widehat{\text{S}}^{(\tmop{off})} |
   \Psi_0 \rangle \right) . \]
Here we presuppose that dispersion-induced spreading does not spoil the
spatial distinctness of the early and the late wave packets. In order to
specify the structure of the scattering matrices
$\widehat{\text{S}}^{(\tmop{on})}$ and $\widehat{\text{S}}^{(\tmop{off})}$, it
is convenient to rewrite the DTE state (\ref{DTEstate}) as
\[ | \Psi_{\tmop{dte}} \rangle = \frac{1}{\sqrt{2}}  \left( |
   \Psi^{(+)}_{\tmop{dte}} \rangle + | \Psi^{(-)}_{\tmop{dte}} \rangle
   \right) \]
with
\[ | \Psi^{(+)}_{\tmop{dte}} \rangle = \frac{1}{\sqrt{2}}  \left(
   \widehat{\text{U}}_{\tau}^{(0)} | \Psi^{(+)}_0 \rangle + \mathe^{\mathi
   \phi_{\tau}} | \Psi^{(+)}_0 \rangle \right), \]
where $| \Psi^{(+)}_0 \rangle = | \psi^{\tmop{cm}}_0 \rangle |
\psi^{\tmop{rel}}_0 \rangle$ and $| \Psi^{(-)}_{\tmop{dte}} \rangle =
\widehat{\text{P}} | \Psi^{(+)}_{\tmop{dte}} \rangle$. Now $|
\Psi^{(+)}_{\tmop{dte}} \rangle$ describes a two-particle state with particle
1 (resp. particle 2) exclusively propagating into positive (negative)
direction. This allows one to assign each particle to a definite interferometer,
e.g., interferometer 1 to particle 1 (and interferometer 2 to particle 2). The
same applies to $| \Psi^{(-)}_{\tmop{dte}} \rangle$, only with the particles
exchanged. Focusing on $| \Psi^{(+)}_{\tmop{dte}} \rangle$, the projection of
the scattered state $| {\Psi^{(+)}_{\tmop{dte}}}' \rangle$ onto a particular
output-port combination $\sigma_1$, $\sigma_2$ reads
\begin{eqnarray}
  \label{transformedDTEstate} \left( \hat{\Pi}_{\sigma_1} \otimes
  \hat{\Pi}_{\sigma_2} \right) | {\Psi^{(+)}_{\tmop{dte}}}' \rangle &=&
  \frac{\widehat{\text{U}}_t^{(0)}}{\sqrt{2}}    \Bigg\{
  \widehat{\text{U}}_{\tau}^{(0)}  \left[
  \widehat{\text{S}}_{\sigma_1}^{(\tmop{on})} \otimes
  \widehat{\text{S}}_{\sigma_2}^{(\tmop{on})}  \right] | \Psi^{(+)}_0 \rangle
\nonumber\\
&&  + \mathe^{\mathi \phi_{\tau}}  \left[
  \widehat{\text{S}}_{\sigma_1}^{(\tmop{off})} \otimes
  \widehat{\text{S}}_{\sigma_2}^{(\tmop{off})} \right] | \Psi^{(+)}_0 \rangle
  \Bigg\} .
\nonumber\\
\end{eqnarray}
Here, $\hat{\Pi}_{\sigma_i}$ is the projection operator onto the region behind
the output port labeled by $\sigma_i = (\pm)_i$ of the {\tmem{i}}th
interferometer. The scattering matrix components $
\widehat{\text{S}}_{\sigma_i}^{(\tmop{on} / \tmop{off})} =
\text{$\hat{\Pi}_{\sigma_i}$}  \widehat{\text{S}}_i^{(\tmop{on} /
\tmop{off})}$ describe the mapping from an {\tmem{in}}-state to the
{\tmem{out}}-state component of a particular beam splitter output port. For
example, $\widehat{\text{S}}_{\sigma_1 = + 1}^{\text{(on)}} | \tmop{in}
\rangle$ yields the {\tmem{out}}-state component in the output port labeled by
$\sigma_1 = + 1$ with the switch in place (``on''). For the early wave
packets, the switch is in place, causing deflection into the long arm. The
offset from the optimum path length difference is reflected in a
translation of the early wave packets with respect to the late ones. The late
wave packets, on the other hand, pass straight through the short arm before
they are distributed into the output ports according to the splitting ratio of
the beam splitter. For the scattering matrix components
$\widehat{\text{S}}_{\sigma_i}^{(\tmop{on})}$ and
$\widehat{\text{S}}_{\sigma_i}^{(\tmop{off})}$, one thus obtains
\begin{eqnarray}
  \widehat{\text{S}}_{\sigma_i = + 1}^{\text{(on)}} & = & \mathe^{\mathi
  \hat{p}_i \ell_i / \hbar} \cos \theta_i,  \label{ScatteringMatrices}\\
  \widehat{\text{S}}_{\sigma_i = - 1}^{\text{(on})} & = & \mathe^{\mathi
  \hat{p}_i \ell_i / \hbar} \sin \theta_i, \nonumber\\
  \widehat{\text{S}}_{\sigma_i = + 1}^{\text{(off)}} & = & \sin \theta_i,
  \nonumber\\
  \widehat{\text{S}}_{\sigma_i = - 1}^{\text{(off})} & = & - \cos \theta_i,
  \nonumber
\end{eqnarray}
where the translation operators $\mathe \tmop{xp} (\mathi \hat{p}_i \ell_i /
\hbar)$ implement the additional displacements $\ell_i$ of the early state
component (switch ``on'') with respect to the late ones. Like for the TBE
state, the joint probability for detecting the particles in a particular
output-port combination $\sigma_1$, $\sigma_2$ is obtained from
$\text{P}_{\tmop{dte}}^{(+)} (\sigma_1, \sigma_2 | \ell_1, \ell_2) = | \left(
\hat{\Pi}_{\sigma_1} \otimes \hat{\Pi}_{\sigma_2}  \right) |
{\Psi^{(+)}_{\tmop{dte}}}' \rangle |^2$. Hence, with
(\ref{transformedDTEstate}) and (\ref{ScatteringMatrices}), we get
\begin{widetext}
\begin{eqnarray*}
  \text{P}_{\tmop{dte}}^{(+)} (\sigma_1, \sigma_2 | \ell_1, \ell_2) & = &
  \frac{1}{4}  \left[ 1 + \sigma_1 \sigma_2 \tmop{Re} \left\{ \mathe^{- \mathi
  \phi_{\tau}} \langle \Psi^{(+)}_0 | \mathe^{\mathi \hat{p}_1 \ell_1 / \hbar}
  \mathe^{- \mathi \hat{p}_1^2 \tau / 2 m \hbar} \otimes \mathe^{\mathi
  \hat{p}_2 \ell_2 / \hbar} \mathe^{- \mathi \hat{p}_2^2 \tau / 2 m \hbar} |
  \Psi^{(+)}_0 \rangle \right\} \right],
\end{eqnarray*}
where we have for simplicity taken the beam splitters to be symmetric
($\theta_i = \pi / 4$) and the particles to be of equal mass $m$. Evaluating
the matrix element in momentum representation, with the abbreviations $\vec{p}
= (p_1, p_2)^{\text{T}}$ and $\text{$\vec{\ell} = (\ell_1,
\ell_2)^{\text{T}}$}$, yields
\begin{eqnarray}
  \text{P}_{\tmop{dte}}^{(+)} (\sigma_1, \sigma_2 | \ell_1, \ell_2) & = &
  \frac{1}{4}  \left[ 1 + \sigma_1 \sigma_2 \tmop{Re} \left\{ \mathe^{- \mathi
  \phi_{\tau}}  \int^{\infty}_{- \infty} \text{d} p_1 \int^{\infty}_{- \infty}
  \text{d} p_2 \mathe^{\mathi \vec{p} \cdot \vec{\ell} / \hbar} \mathe^{-
  \mathi \vec{p}^2 \tau / 2 m \hbar} | \langle p_1, p_2 | \Psi^{(+)}_0 \rangle
  |^2 \right\} \right] . \nonumber\\
  &  &  \label{DTEdetectionProbability}\\
  &  &  \nonumber
\end{eqnarray}
\end{widetext}
The intermediate result (\ref{DTEdetectionProbability}) already reveals some
important features of the setup: Firstly, the overall free time evolution
$\widehat{\text{U}}_t^{(0)}$ in (\ref{transformedDTEstate}) drops out for the
detection probability, as it was the case for the TBE state. Hence, only the
dispersion-induced distortion between the early and late wave packets, which
is due to the period $\tau$ between the two dissociation processes, causes
potential harm to the fringe pattern of the detection probability as a
function of the arm length variations. Next, only the momentum distribution $|
\langle p_1, p_2 | \Psi^{(+)}_0 \rangle |^2$ of the single dissociation pulse
component $| \Psi^{(+)}_0 \rangle$ enters the detection probability
$\text{P}_{\tmop{dte}}^{(+)} (\sigma_1, \sigma_2 | \ell_1, \ell_2)$. As a
consequence, $\text{P}_{\tmop{dte}}^{(+)} (\sigma_1, \sigma_2 | \ell_1,
\ell_2)$ is invariant under momentum phase transformations $\langle p_1, p_2 |
\Psi_0 \rangle \rightarrow \exp [\mathi \xi (p_1, p_2)] \langle p_1, p_2 |
\Psi_0 \rangle$, which includes spatial translations. This means that the
signal is unaffected by shot-to-shot shifts of the source position with
respect to the interferometers (as long as the interferometric procedure is
still feasible). The mere dependence on the momentum distribution also shows
that it is straightforward to generalize the detection probability to nonpure
and nonseparable states, $| \Psi^{(+)}_0 \rangle \langle \Psi^{(+)}_0 |
\rightarrow \rho^{(+)}_0$, and correspondingly $| \langle p_1, p_2 |
\Psi^{(+)}_0 \rangle |^2 \rightarrow \tmop{pr} (p_1, p_2) \assign \langle p_1,
p_2 | \rho^{(+)}_0 |p_1, p_2 \rangle$.

In the end, we are interested in the functional dependence of the detection
probability on the variations $\ell_i$ of the interferometer arm lengths. In
order to get a qualitative and quantitative understanding of the resulting
fringe pattern, we evaluate Eq. (\ref{DTEdetectionProbability}) in closed form
for the case that the early and the late wave packet components are each
described by a generic (mixed) Gaussian state. This is reasonable, since even
for non-Gaussian wave packet components with a more complicated momentum
distribution, an appropriate Gaussian fit should allow us to derive at least a
lower bound to the quality of the fringe pattern. For a Gaussian momentum
distribution,
\begin{widetext}
\[ \tmop{pr} (p_1, p_2) = \frac{1}{2 \pi \sigma_{p, \tmop{cm}} \sigma_{p,
   \tmop{rel}}} \exp \left( - \frac{(p_1 + p_2)^2}{2 \sigma^2_{p, \tmop{cm}}}
   - \frac{(p_1 - p_2 - m v_{\tmop{rel}})^2}{8 \sigma^2_{p, \tmop{rel}}}
   \right), \]
we arrive at
\begin{eqnarray}
  \text{P}_{\tmop{dte}}^{(+)} (\sigma_1, \sigma_2 | \ell_1, \ell_2) & = &
  \frac{1}{4}  \left\{ 1 + \sigma_1 \sigma_2  \left( 1 +
  \frac{\tau^2}{T^2_{\tmop{cm}}} \right)^{- 1 / 4}  \left( 1 +
  \frac{\tau^2}{T^2_{\tmop{rel}}} \right)^{- 1 / 4} 
  \label{DTEGaussianDetectionProbability} \right.\\
  &  & \times \mathe \tmop{xp} \left[ -
  \frac{T_{\tmop{rel}}}{T^2_{\tmop{rel}} + \tau^2}  \frac{(\ell_1 - \ell_2 -
  \tau v_{\tmop{rel}})^2}{2 v_{\tmop{rel}} \lambdabar_{\text{rel}}} -
  \frac{T_{\tmop{cm}}}{T^2_{\tmop{cm}} + \tau^2}  \frac{(\ell_1 + \ell_2)^2}{2
  v_{\tmop{rel}} \lambdabar_{\text{rel}}} \right] \nonumber\\
  &  & \left. \times \cos \left[ \frac{\ell_1 - \ell_2}{\lambdabar_{\text{rel}}} +
  \frac{\tau}{T^2_{\tmop{rel}} + \tau^2}  \frac{(\ell_1 - \ell_2 - \tau
  v_{\tmop{rel}})^2}{2 v_{\tmop{rel}} \lambdabar_{\text{rel}}} + \frac{\tau}{T^2_{\tmop{cm}} +
  \tau^2}  \frac{(\ell_1 + \ell_2)^2}{2 v_{\tmop{rel}} \lambdabar_{\text{rel}}} -
  \frac{\varphi_0}{2} \right] \right\} \nonumber
\end{eqnarray}
\end{widetext}
with $\varphi_0 = \tau v_{\tmop{rel}} / \lambdabar_{\text{rel}} + \arctan (\tau /
T_{\tmop{cm}}) + \arctan (\tau / T_{\tmop{rel}}) + 2 \phi_{\tau}$. The
variances of the relative and the center-of-mass momentum, denoted by
$\sigma_{p, \tmop{rel}}$ and $\sigma_{p, \tmop{cm}}$, respectively, determine
characteristic dispersion times, $T_{\tmop{cm}} = 2 m \hbar / \sigma^2_{p,
\tmop{cm}}$ and $T_{\tmop{rel}} = m \hbar / 2 \sigma^2_{p, \tmop{rel}}$. The
latter indicate the time scale of transition to a dispersion-dominated spatial
extension of the wave packets. The expectation value of the relative momentum
$p_{0, \tmop{rel}} = m v_{\tmop{rel}} / 2$ defines the reduced wave length
$\lambdabar_{\text{rel}} = \hbar / p_{0, \tmop{rel}}$, which sets the scale for the nonlocal
interference fringes.

The DTE detection probability (\ref{DTEGaussianDetectionProbability}) has the
potential to violate a Bell inequality, as is seen from its structural
similarity to the TBE detection probability (\ref{TBEdetectionProbability}).
The dispersion-induced distortion between the early and the late wave packet
components is reflected in those terms in
(\ref{DTEGaussianDetectionProbability}) which depend on the characteristic
dispersion times $T_{\tmop{cm}}$ and $T_{\tmop{rel}}$. In particular, it is
responsible for the overall suppression of the fringe pattern as described by
the Lorentzian reduction factors. Further, it causes a quadratic compression
of the fringe pattern. The additional Gaussian suppression is due to the
unavoidable envelope mismatch that follows from the variation of the arm
lengths.

In order to get an unambigous demonstration of nonlocal correlations, the
fringe visibility has to exceed the threshold value $1 / \sqrt{2}$ over at
least a few fringes when varying either arm length by $\ell_i$. An analysis of
the joint detection probability (\ref{DTEGaussianDetectionProbability}) shows
that this is achieved given the following two conditions are met:
\begin{eqnarray}
  \lambdabar_{\text{rel}} / (\tau v_{\tmop{rel}}) \ll 1, &  &  \label{CompressionCondition}\\
  (1 + \tau^2 / T^2_{\tmop{cm}}) (1 + \tau^2 / T^2_{\tmop{rel}}) < 4. &  & 
  \label{VisibilityCondition}
\end{eqnarray}
So far, we have restricted our investigation to the DTE state $|
\Psi^{(+)}_{\tmop{dte}} \rangle$ with each particle propagating into a given
direction. It is clear that we could have followed the same reasoning for the
DTE state $\text{$| \Psi^{(-)}_{\tmop{dte}} \rangle$}$, only with the labels
for the particles exchanged, so that the corresponding joint detection
probability $\text{P}_{\tmop{dte}}^{(-)} (\sigma_1, \sigma_2 | \ell_1,
\ell_2)$ equals $\text{P}_{\tmop{dte}}^{(+)} (\sigma_1, \sigma_2 | \ell_1,
\ell_2)$. Finally, since no interference between $\text{$|
\Psi^{(+)}_{\tmop{dte}} \rangle$}$ and $\text{$| \Psi^{(-)}_{\tmop{dte}}
\rangle$}$ occurs in our setup, the detection probability for the symmetric
DTE state $\text{$| \Psi_{\tmop{dte}} \rangle = \frac{1}{\sqrt{2}}  \left( |
\Psi^{(+)}_{\tmop{dte}} \rangle + | \Psi^{(-)}_{\tmop{dte}} \rangle \right)$}$
follows from
\begin{eqnarray*}
  \text{P}_{\tmop{dte}}^{} (\sigma_1, \sigma_2 | \ell_1, \ell_2) & = &
  \frac{1}{2}  \Big( \text{P}_{\tmop{dte}}^{(+)} (\sigma_1, \sigma_2 |
  \ell_1, \ell_2)\\
&&\phantom{\frac{1}{2}  \Big(} + \text{P}_{\tmop{dte}}^{(-)} (\sigma_1, \sigma_2 | \ell_1,
  \ell_2) \Big)\\
  & = & \text{P}_{\tmop{dte}}^{(+)} (\sigma_1, \sigma_2 | \ell_1, \ell_2) .
\end{eqnarray*}
Hence, the results obtained for the directed DTE state $|
\Psi^{(+)}_{\tmop{dte}} \rangle$ apply just as well to the experimentally
realized DTE state $| \Psi_{\tmop{dte}} \rangle$.

We thus find that an unambigous verification of nonlocal correlations by
violating a Bell inequality can be achieved in the setup considered above,
given the conditions (\ref{CompressionCondition}) and
(\ref{VisibilityCondition}) are satisfied. A recent proposal is based on the
Feshbach-induced dissociation of a molecular Bose--Einstein condensate
{\cite{Gneiting2008a}}. As is argued there, an experiment based on fermionic
Lithium atoms would indeed meet the above conditions
(\ref{CompressionCondition}) and (\ref{VisibilityCondition}), and hence raises
hope for demonstrating nonclassicality in the motion of material particles.
This setup would involve time separations between the two dissociation pulses
on the order of seconds. This would result in a spatial separation between the
early and late wave packets on the order of centimeters, rendering the DTE
state truly macroscopic. \

\section{\label{Conclusions}Conclusions}

We presented a scheme to generate and verify nonlocal correlations between
two material particles involving macroscopic superpositions of the spatial
wave function. This is achieved by violating a Bell inequality using
single-particle matter wave optics and simple position measurements. We
conclude by summarizing the advantages of the presented scheme. First of all,
it does not require interferometric stability between the two interferometers,
which allows one to reach truly macroscopic separations. Moreover, the restriction
to the coherence properties between the early and late wave packets as a whole
means that only port-selective measurements are needed, requiring neither
prominent spatial nor temporal resolution. This renders the nonclassical
correlations largely independent of the shape of the wave packets and of the
particular implementation of the position measurement. Finally, the low
velocity of the propagating atoms in principle allows one to check the particle
positions by laser illumination whenever one wishes. One might thus think of
a demonstration experiment, where only after the dissociation of the particles
the experimenter makes the conscious decision whether she wants to check the
correlation of the emission times (by ``looking'' at the particles in front of
the interferometers) or whether she wants to check the nonlocal correlations
(by ``looking'' at the particles behind of the interferometers). This way, by
performing such an experiment with proper, material particles and on truly
macroscopic scales on the order of centimeters, nonclassical quantum
correlations would be made amenable to anyone who understands the basic
concept of the position of a particle, even to a layman who is ignorant of
physics. All this renders the proposed setup a rather robust and striking test
of nonclassicality in the motion of material particles with the potential to
push the corresponding quantum regime into the macroscopic.

As an outlook, it would be illuminating to investigate the influence of
various sources of decoherence. For the experimental proposal in
{\cite{Gneiting2008a}}, the effect of scattering of off-resonant photons and
the scattering of background particles was already found to be controllable.
In particular, it would be interesting to check to what degree the present
setup can be used to test possible unconventional collapse theories, which
predict a loss of coherence in the motion of material particles due to
purported quantum gravity effects, on the centimeter scale accessible for the
first time with the present measurement scheme.


\begin{thebibliography}{10}
  \bibitem[1]{Einstein1935a}A.~Einstein, B.~Podolsky, and N.~Rosen,
  {\newblock}Phys. Rev. \tmtextbf{47}, 777 (1935).
  
  \bibitem[2]{Schrodinger1935a}E.~Schr\"odinger, {\newblock}Proc. Camb. Phil.
  Soc. \tmtextbf{31}, 555 (1935).
  
  \bibitem[3]{Bell1964a}J.~S. Bell, {\newblock}Physics \tmtextbf{1}, 195
  (1964).
  
  \bibitem[4]{Aspect1982a}A.~Aspect, J.~Dalibard, and G.~Roger,
  {\newblock}Phys. Rev. Lett. \tmtextbf{49}, 1804 (1982).
  
  \bibitem[5]{Gneiting2008a}C.~Gneiting and K.~Hornberger,   {\newblock}Phys. Rev. Lett. \tmtextbf{101}, 260503 (2008).
  
  \bibitem[6]{Kheruntsyan2005a}K.~V. Kheruntsyan, M.~K. Olsen, and P.~D.
  Drummond, {\newblock}Phys. Rev. Lett. \tmtextbf{95}, 150405 (2005).
  
  \bibitem[7]{Opatrny2001a}T.~Opatrn\'y and G.~Kurizki, {\newblock}Phys. Rev.
  Lett. \tmtextbf{86}, 3180 (2001).
  
  \bibitem[8]{Banaszek1998a}K.~Banaszek and K.~Wodkiewicz, {\newblock}Phys.
  Rev. A \tmtextbf{58}, 4345 (1998).
  
  \bibitem[9]{Chen2002a}Z.-B. Chen, J.-W. Pan, G.~Hou, and Y.-D. Zhang,
  {\newblock}Phys. Rev. Lett. \tmtextbf{88}, 040406 (2002).
  
  \bibitem[10]{Brendel1999a}J.~Brendel, N.~Gisin, W.~Tittel, and H.~Zbinden,
  {\newblock}Phys. Rev. Lett. \tmtextbf{82}, 2594 (1999).
  
  \bibitem[11]{Tittel2000a}W.~Tittel, J.~Brendel, H.~Zbinden, and N.~Gisin,
  {\newblock}Phys. Rev. Lett. \tmtextbf{84}, 4737 (2000).
  
  \bibitem[12]{Simon2005a}C.~Simon and J.~P. Poizat, {\newblock}Phys. Rev.
  Lett. \tmtextbf{94}, 030502 (2005).
  
  \bibitem[13]{Franson1989a}J.~D. Franson, {\newblock}Phys. Rev. Lett.
  \tmtextbf{62}, 2205 (1989).
  
  \bibitem[14]{Mukaiyama2004a}T.~Mukaiyama, J.~Abo-Shaeer, K.~Xu, J.~Chin, and
  W.~Ketterle, {\newblock}Phys. Rev. Lett. \tmtextbf{92}, 180402 (2004).
  
  \bibitem[15]{Durr2004b}S.~D\"urr, T.~Volz, and G.~Rempe, {\newblock}Phys.
  Rev. A \tmtextbf{70}, 031601 (2004).
  
  \bibitem[16]{Allcock1969a}G.~R. Allcock, {\newblock}Ann. Phys.
  \tmtextbf{53}, 253 (1969).
  
  \bibitem[17]{Werner1986a}R.~Werner, {\newblock}J. Math. Phys. \tmtextbf{27},
  793 (1986).
  
  \bibitem[18]{Aharonov1998a}Y.~Aharonov, J.~Oppenheim, S.~Popescu, B.~Reznik,
  and W.~G. Unruh, {\newblock}Phys. Rev. A \tmtextbf{57}, 4130 (1998).
  
  \bibitem[19]{Muga2000a}J.~G. Muga and C.~R. Leavens, {\newblock}Phys. Rep.
  \tmtextbf{338}, 353 (2000).
  
  \bibitem[20]{Clauser1969a}J.~F. Clauser, M.~A. Horne, A.~Shimony, and R.~A.
  Holt, {\newblock}Phys. Rev. Lett. \tmtextbf{23}, 880 (1969).
  
  \bibitem[21]{Aerts1999a}S.~Aerts, P.~Kwiat, J.-A. Larsson, and M.~{\.Z}ukowski,
  {\newblock}Phys. Rev. Lett. \tmtextbf{83}, 2872 (1999).
  
  \bibitem[22]{Cabello2008a}A.~Cabello, A.~Rossi, G.~Vallone, F.~De~Martini,
  and P.~Mataloni, {\newblock}Phys. Rev. Lett. \tmtextbf{102}, 040401 (2009).
  
  \bibitem[23]{Taylor1972a}J.~R.~Taylor,
  {\newblock}\tmtextit{Scattering Theory: The Quantum Theory on
  Nonrelativistic Collisions} (Wiley, New York, 1972).
\end{thebibliography}
\end{document}